# DeepPNI: Language- and graph-based model for mutation-driven protein-nucleic acid energetics

Somnath Mondal[1], Tinkal Mondal[2], Soumajit Pramanik[3], Rukmankesh Mehra[1,2,*]

[1]*Department of Chemistry, Indian Institute of Technology Bhilai, Durg 491002, Chhattisgarh, India.*

[2]*Department of Bioscience and Biomedical Engineering, Indian Institute of Technology Bhilai, Durg 491002, Chhattisgarh, India.*

[3]*Department of Computer Science and Engineering, Indian Institute of Technology Bhilai, Durg-491002, Chhattisgarh, India.*

[*]Correspondence: rukmankesh@iitbhilai.ac.in

ORCID
*Somnath Mondal: 0009-0002-6997-5968*
*Tinkal Mondal: 0000-0003-0840-2664*
*Soumajit Pramanik: 0000-0002-2297-1153*
*Rukmankesh Mehra: 0000-0001-6010-1514*

# Abstract

The interaction between proteins and nucleic acids is crucial for processes that sustain cellular function, including DNA maintenance and the regulation of gene expression and translation. Amino acid mutations in protein-nucleic acid complexes often lead to vital diseases. Experimental techniques have their own specific limitations in predicting mutational effects in protein-nucleic acid complexes. In this study, we compiled a large dataset of 1951 mutations including both protein-DNA and protein-RNA complexes and integrated structural and sequential features to build a deep learning-based regression model named DeepPNI. This model estimates mutation-induced binding free energy changes in protein-nucleic acid complexes. The structural features are encoded via edge-aware RGCN and the sequential features are extracted using protein language model ESM-2. We have achieved a high average Pearson correlation coefficient (PCC) of 0.76 in the large dataset via five-fold cross-validation. Consistent performance across individual dataset of protein-DNA, protein-RNA complexes, and different experimental temperature split dataset make the model generalizable. Our model showed good performance in complex-based five-fold cross-validation, which proved its robustness. In addition, DeepPNI outperformed in external dataset validation, and comparison with existing tools.

# 1. Introduction

Cellular processes are maintained through communication among numerous biomolecules. Protein-nucleic acid interactions (PNIs) play key roles in the regulation of essential biological functions such as DNA replication, repair, recombination, gene expression, and translation.[1,2] The affinity between proteins and nucleic acids (DNA and RNA) within a protein-nucleic acid complex is governed by intermolecular forces, physicochemical properties, and structural features.[3,4] These factors are often disrupted by missense mutations occurring in either or both counterparts of the complex.[3]

Numerous pathologies related to these mutations have been described in the literature.[5–16] For instance, mutations in TDP-43 (Transactive response DNA-binding protein 43), an RNA/DNA-binding protein, have been shown to be linked with neurodegenerative diseases like ALS (amyotrophic lateral sclerosis) and FTD (frontotemporal dementia).[6] Melanoma, a type of fatal skin cancer, is mainly caused by solar-radiation induced mutations in DNA.[7] It has been observed that these DNA mutations can amplify TERT (Telomerase) promoter activity. In normal conditions, TERT in complex with RNA contributes to the synthesis of telomeres at chromosomal ends. However, its abnormally heightened activity can promote carcinogenesis.[5] Further, Leigh syndrome is a progressive neurodegenerative disorder that has been reported to be caused by pathogenic mutations in mitochondrial tRNAs. Upon mutation, these tRNAs become substrates for different aminoacyl-tRNA synthetases thus hindering the usual protein synthesis.[17–19] These examples show that mutations in protein-nucleic acid complexes trigger several diseases, and therefore the effects of these alterations are crucial to be investigated thoroughly.

The impact of a mutation can be assessed experimentally using techniques such as surface plasmon resonance, FRET, or isothermal titration calorimetry to determine the change in binding affinity.[20–24] However, with the rapid growth of genomic data, the demand for high-throughput analysis is also increasing proportionally.[25] Conventional experimental methods are not well suited for this purpose due to their slow throughput and high expense.[26] The advent of high-throughput experimental methods such as high-throughput SELEX,[27] protein-binding microarrays,[28] mechanically induced trapping of molecular interactions[29] has partially addressed this challenge. Nevertheless, these techniques also have limitations restricting their wide applicability.[28–30] Computational methods such as free energy perturbation and thermodynamic integration can accurately calculate binding free energy.[31] However, these methods are also not appropriate for large scale studies because they are computationally expensive.[32–37]

In this work, we developed a deep learning model for PNI predictions trained on a larger dataset of 1951 mutations across protein-DNA and protein-RNA complexes from NABE[38] database. The structural and sequential features were fused together, and multiple types of nodes and their edges were used to generate an edge-aware relational graph convolutional network to encode the structural features. These structural features were integrated with the sequential features from protein language model to predict the binding free energy change upon amino acid mutation. We tested our model's performance on a large variety of data splits that included individual protein-DNA, protein-RNA data, and temperature-based split. The robustness of our model was confirmed by its good performance in complex-based five-fold cross-validation experiment. Our model outperformed the recent state-of-the-art methods and showed a good performance in external database validation. The combination of sequential and edge-aware atomic level relational graph convolutional network in a large dataset made this work innovative.

## 2. Methods

### 2.1. Study design

This study aims to quantitatively predict how single amino acid mutations across protein-nucleic acid interface affect their binding affinity. The impact of a mutation is measured by the binding free energy change ($\Delta\Delta G$), defined as the difference in binding free energies between the mutant and the wild-type complex:

$$\Delta\Delta G = \Delta G_{mutant} - \Delta G_{wild\text{-}type}$$

A positive $\Delta\Delta G$ indicates a destabilizing mutation (reduced binding affinity), whereas a negative value represents a stabilizing mutation (enhanced binding affinity).

This task is formulated as a supervised regression problem, where each mutation is represented by a combination of structural and sequential features. Specifically, a local atomic graph centered at the mutation site was encoded using a graph convolutional network (GCN) to capture spatial and physicochemical information, while an ESM-based protein language model embedding described the sequence context. The concatenated feature representation $z_{combined} = [z_G \parallel e_{esm}]$ was used to learn a mapping function $f: z_{combined} \rightarrow \Delta\Delta G$, predicting the experimentally measured change in binding free energy upon mutation.

**Table 1. Dataset summary**

| Dataset | Number of mutations | Description |
| --- | --- | --- |
| PN | 1951 | Protein-nucleic acid complex |
| PD | 1328 | Protein-DNA complex |
| PR | 620 | Protein-RNA complex |
| PDR-hc | 3 | Protein-DNA-RNA hybrid complex |

## 2.2. Dataset construction

To build a regression model for predicting the amino acid mutational effects at protein-nucleic acid interfaces, we collected data from the NABE[38] database, which contains experimentally measured changes in binding affinity for over 2,500 mutations across more than 400 protein-DNA and protein-RNA complexes. Since multiple experimental records existed for several mutations, duplicate entries were averaged to ensure consistency. The resulting curated dataset comprised a total of 1,951 unique mutations, including 1,328 from protein-DNA complexes, 620 from protein-RNA complexes (**Figure 3a** and **Table 1**), and three from a protein-DNA-RNA hybrid complex (PDB ID: 4WB2). For clarity, we labeled the complete dataset as PN (protein–nucleic acid), while the subsets corresponding to protein-DNA and protein-RNA complexes were labeled as PD and PR, respectively. Based on the experimental temperature conditions, the dataset was further divided into three groups: @298 K, @>298 K, and @<298 K.

For external validation, we used the ProNAB[39] database, which provides experimental data on protein-nucleic acid interactions. The ProNAB dataset was categorized according to the experimental technique employed (e.g., ITC, FB, FP, GS), as detailed in **Supplementary Section 1**. In addition, we utilized two benchmark datasets MPR311 previously reported by Dong-Jun Yu et al.,[40] and PremPDI[41] to further assess the generalizability of our model.

## 2.3. Sequential feature generation

To incorporate sequence-level information, we extracted protein sequence contexts surrounding each mutation directly from the corresponding PDB[42] structures. For each mutation, the wild-type residue and its positional index were parsed and validated against the structural chain to ensure consistency between sequence annotation and 3D coordinates. A continuous amino acid sequence was reconstructed for each chain, accounting for missing residues by inserting a padding character ('X'). A local sequence window of ±200 residues centered on the mutation site was selected to capture the mutational environment.

Each extracted sequence was then embedded using pre-trained protein language model: ESM-2[43] with 1280-dimensional embedding. This embedding was obtained via mean pooling across the sequence length, yielding fixed length numerical representations of the local mutation context. These embedding encode higher order structural and evolutionary features learned from large-scale protein corpora, providing complementary sequential information to the structural graph representation used for $\Delta\Delta G$ regression modelling.

## 2.4. Graph-based structural feature engineering

Each mutational interface of protein-nucleic acid complex was represented as an undirected atomic graph $G = (V, E, T_v, T_e)$, where $V$ and $E$ denote the sets of nodes (atoms) and edges (atomic interactions), respectively (**Figure 1**). There is an initial feature matrix $X \in \mathbb{R}^{|V| \times d}$ with $d = 12$ atomic feature per node, $T_v$ is the node type assignment, and $T_e$ is the edge type assignment. Three-dimensional atomic coordinates were obtained from PDB structures. For each mutation site, a local environment was defined by including all atoms within a sphere of 10 Å radius from the geometric centre of wild-type target residue.

Each node $v_i \in V$ was characterized by a 12-dimensional feature vector $x_i$ encoding fundamental atomic properties: atomic number, degree, formal charge, aromaticity, implicit hydrogen count, and a seven-dimensional one-hot encoding of hybridization type (sp, $sp^2$, $sp^3$, $sp^3d$, $sp^3d^2$, s, and unspecified). To incorporate biological context, nodes were classified into three distinct types: wild-type residue atoms ($t_v = 0$), other protein residue atoms ($t_v = 1$), and nucleic acid atoms ($t_v = 2$). This heterogeneous node typing enabled the model to distinguish mutation sites from their structural environment.

Edges $e_{ij} \in E$ between nodes $v_i$ and $v_j$ were established based on inter-atomic distances and interaction types. Edges were created between atom pairs within 4 Å to capture spatial connectivity and according to all covalent edges between atoms. Six different types of edges ($t_e \in \{0,1,2,3,4,5\}$) were also labeled as: edge between atoms of wild-type residue ($e_{ww}$), edge between atoms of wild-type residue and other residue of protein ($e_{wo}$), edge between wild-type residue atom and nucleic acid atom ($e_{wn}$), edge between atoms of other residues ($e_{oo}$), edge between other residue and nucleic acid atom ($e_{on}$), edge between atoms of nucleic acid residue ($e_{nn}$).

The resulting graphs were stored with the node feature matrix $X$, edge $E$, and their different relation. Each graph was labeled with the experimentally measured binding free energy change ($\Delta\Delta G = \Delta G_{mutant} - \Delta G_{wild-type}$) corresponding to the mutation. These graph representations formed the input for downstream regression task through edge aware relational graph convolutional network (RGCN) to predict the mutational effects on protein-nucleic acid binding affinity.

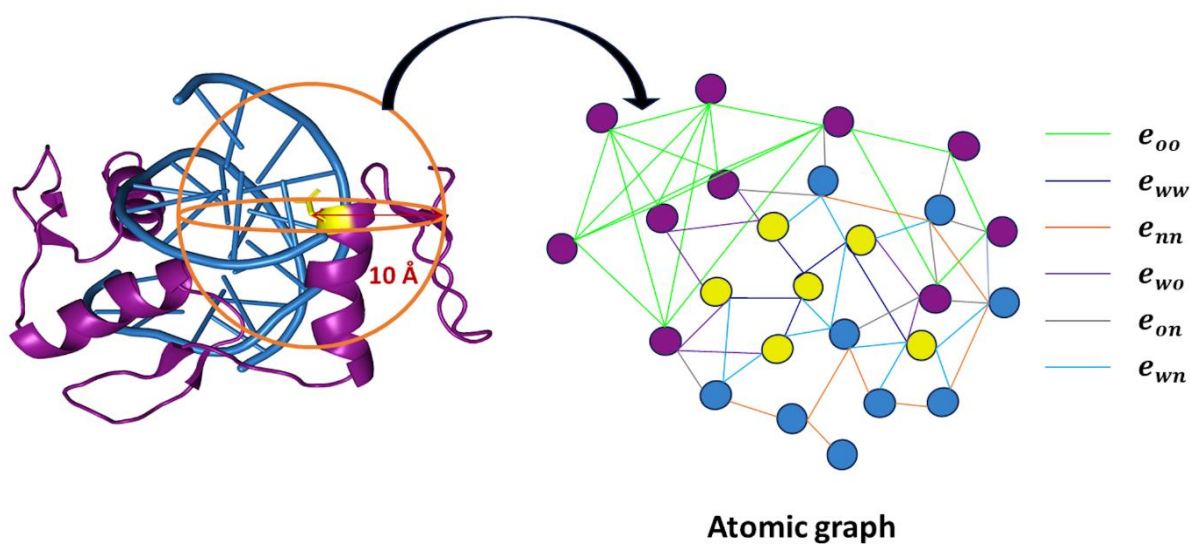

**Figure 1. Atomic graph within 10 angstrom spheres from geometric centre of wild-type residue.** Wild-type residue atoms, other protein residue atoms, and nucleic acid atoms are coloured as yellow, purple, and blue. Six different edges ( between atoms of wild-type residue ($e_{ww}$), edge between atoms of wild-type residue and other residue of protein ($e_{wo}$), edge between wild-type residue atom and nucleic acid atom ($e_{wn}$), edge between atoms of other residues ($e_{oo}$), edge between other residue and nucleic acid atom ($e_{on}$), edge between atoms of nucleic acid residue ($e_{nn}$) are coloured according to the labels.

## 2.5. Heterogeneous graph embedding generation

Heterogeneous graph level embedding generation consisted of three components, input feature transformation, stacked edge-aware relational graph convolutional network (RGCN) layers with residual connections, and a graph level pooling layer. This architecture extracted the complex structural heterogenicity of protein-nucleic acid complex around the mutation site.

The initial atomic features $x_i$ were transformed to a higher dimensional space to enable richer representations. An input projection layer mapped node features to a hidden dimension of 256 via

$$h_i^{(0)} = Dropout(ReLU(BN(W_{in}x_i + b_{in})))$$

where, $W_{in} \in \mathbb{R}^{256 \times 12}$ is the projection weight matrix, BN is the batch normalization. And dropout rate of 0.1 was applied for regularization. This linear transformation created an initial node representation $h_i^{(0)} \in \mathbb{R}^{256}$ for each node.

Beyond traditional RGCN, we used an edge-aware mechanism where individual edge was treated with different importance. This was done through two parallel learned components within each RGCN layer, firstly edge-specific weight learning and secondly relation-specific feature transformation. For each type of $r \in \{0,1,2,3,4,5\}$, model learned a low dimensional edge embedding $e_r = EdgeEmb(r) \in \mathbb{R}^{16}$. These embeddings were initialized randomly and trained end-to-end. For each edge of type $t_e$, we calculated edge specific attention weight using a two-layer neural network.

$$w_{ij} = \sigma(W_2^{edge} \cdot ReLU(W_1^{edge} e_{t_e} + b_1^{edge}) + b_2^{edge})$$

where, $W_1^{edge} \in \mathbb{R}^{16 \times 16}$, $W_2^{edge} \in \mathbb{R}^{1 \times 16}$, and $\sigma$ is the sigmoid activation function. This learned weight allowed the model to rely on the important edges in the message propagation mechanism. Now, each edge type processed the node features through its own relation specific transformation matrix. We employed basis decomposition to manage the parameter complexity. Here, instead of full weight matrix $W_r \in \mathbb{R}^{256 \times 256}$ for each relation type $r$, we decomposed them as $W_r^{(l)} = \sum_{b=1}^{B} a_{rb}^{(l)} V_b^{(l)}$ where, $V_b^{(l)} \in \mathbb{R}^{256 \times 256}$ were $B = 3$ shared basis matrices, and $a_{rb}^{(l)}$ were relation-specific coefficients. This reduced the parameters from $6 \times 256^2$ to $(3 \times 256^2 + 3 \times 6)$.

Now edge-specific learned weights and relation-specific transformations jointly completed the message passing operation in layer $l$. The aggregated message from all the neighbours for $i$ node was computed as

$$m_i^{(l)} = \sum_{r=0}^{5} \sum_{j \in \mathcal{N}_i^r} \frac{w_{ij}}{\sqrt{|\mathcal{N}_i^r \| \mathcal{N}_j^r|}} W_r^{(l)} h_j^{(l)}$$

where, $\mathcal{N}_i^r$ denoted the set of neighbour node $i$ connected by edge of type $r$, $w_{ij}$ and $W_r^{(l)}$ were the edge-specific weight and relation-specific transformation matrix, the normalization factor $\sqrt{|\mathcal{N}_i^r \| \mathcal{N}_j^r|}$ prevented the gradient instability in regions with varying node degrees.

Here the main interesting part was the message passing from node $j$ to $i$, which was computed as, $Message_{j \to i} = w_{ij} \times (W_{t_e} h_j^{(l)})$. In this formulation the main intuition was, $W_{t_e} h_j^{(l)}$ transformed the neighbour's features according to the relationship type, and $w_{ij}$ scaled this transformed message based on the learned importance of this specific edge instance. The node update also included a self-connection term to retain information from the previous layer,

$$h_i^{(l+1)} = ReLU(m_i^l + W_0^{(l)} h_i^{(l)} + b^{(l)})$$

where, $W_0^{(l)} \in \mathbb{R}^{256 \times 256}$ was the self-connection weight matrix, $b^{(l)}$ was a bias term.

The complete graph embedding generation consisted of four RGCN layers with batch normalisation, ReLU activation, dropout, and residual connection. This deep architecture enabled hierarchical feature learning for graph representation by capturing local atomic interactions. The layer wise transformation could be written as

$$h_i^{(l+1)} = h_i^{(l)} + Dropout(ReLU\left(BN\left(RGCN^{(l)}\left(h_i^{(l)}, G\right)\right)\right))$$

where, $RGCN^{(l)}$ represented the complete edge-aware message passing operation. The residual connection (adding $h_i^{(l)}$) helped to mitigate the vanishing gradient problem in this network and enabled better gradient flow during training. After the RGCN block each node had 256 dimensional embeddings. We used mean pooling technique to get the graph level embedding. We projected the 256-dimensional graph embedding to 128-dimensional space and used the lower dimensional vector $z_G$ in the subsequent fusion stage.

## 2.6. Dual feature integration and fusion network

The structural information encoded in the GCN embedding was combined with sequential embeddings from protein language model to create comprehensive representation of the mutation context. The integration strategy employed feature concatenation to preserve all modality specific information:

$$z_{combined} = [z_G \parallel e_{esm}]$$

where $z_G \in \mathbb{R}^{128}$ is the GCN graph embedding capturing structural topology, $e_{esm} \in \mathbb{R}^{1280}$ is the ESM-2 embedding capturing sequence-structure relationships. The concatenated vector $z_{combined} \in \mathbb{R}^{d_{total}}$ has the total dimensionality $d_{total} = 128 + 1280 = 1408$.

This dual representation was subsequently processed through a deep multi-layer perception (MLP) to predict $\Delta\Delta G$ value. The MLP architecture consisted of three hidden layers [512, 256, 64] with progressively decreasing dimensionality, implementing a funnel like structure that gradually distils the high dimensional feature into a scaler prediction. Batch normalisation at each MLP layers ensures stable gradients and faster convergence, while dropout with a probability $p = 0.3$ prevents overfitting to the training data. The final layer produced a scalar output representing the predicted change in binding free energy ($\Delta\Delta G$) in kcal/mol, without activation to allow predictions across the full real number range.

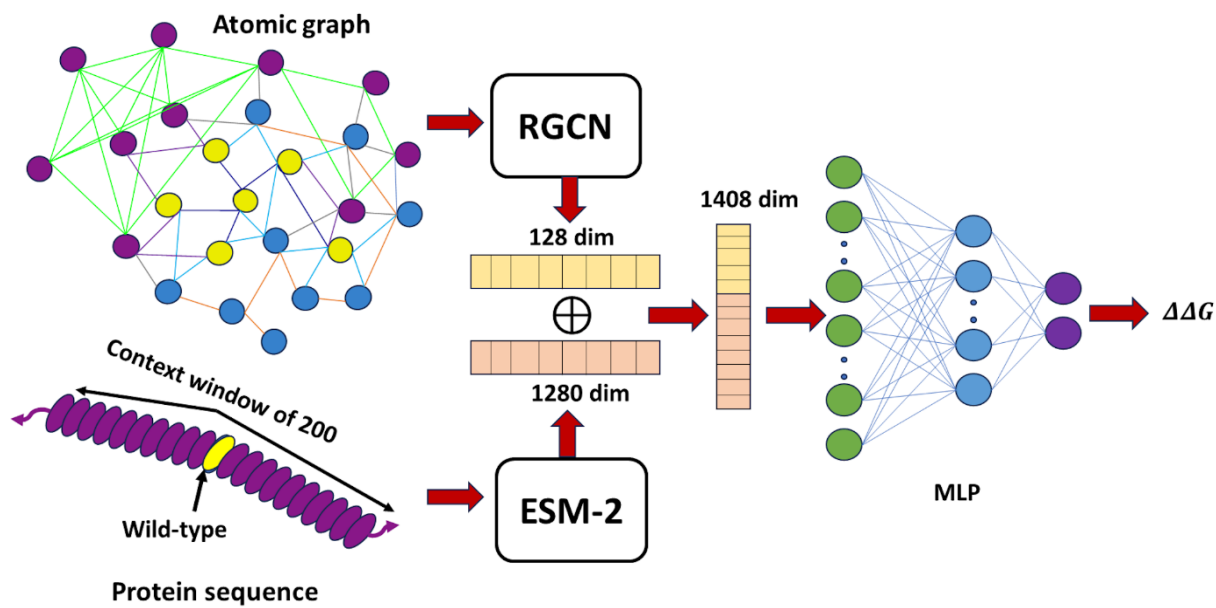

**Figure 2. Architecture of DeepPNI, integrating graph convolutional and ESM-2 embeddings for $\Delta\Delta G$ prediction.** Graph level embedding (structural information) is from edge-aware RGCN and sequential information is from ESM-2.

## 2.7. Training procedure

The GNN encoder and MLP predictor were jointly trained end-to-end, while the protein language model embeddings were pre-computed and frozen. The complete model optimized the following objective:

$$\mathcal{L} = \frac{1}{N}\sum_{i=1}^{N}(\hat{y}_i - y_i)^2 + \lambda \, \|\ominus\|_2^2$$

where $N$ is the batch size, $y_i$ is the experimental $\varDelta\varDelta G$ value, $\hat{y}_i$ is the prediction, $\ominus$ represents all trainable parameters, and $\lambda = 10^{-4}$ is the L2 regularization coefficient. Optimization was performed using Adam optimizer with learning rate $\eta = 0.0001$. Gradient clipping with maximum norm 1.0 was applied to prevent exploding gradients. The learning rate was dynamically adjusted using ReduceLROnPlateau scheduling.

Five-fold cross validation was implemented to ensure robust performance estimation. For each fold, the model was trained for 250 epochs with a batch size 32. The model checkpoint with the lowest validation loss was retained for evaluation. To enhance model robustness, we implemented z-score outlier detection with a threshold value of 1.5. The performance of our model was estimated based on PCC (Pearson correlation coefficient), RMSE (root mean squared error), and MAE (mean absolute error).

## 2.8. Implementation

All models were implemented in PyTorch 2.0+ utilizing PyTorch Geometric for graph operation. Training was performed on NVIDIA RTX A5000 (CUDA 11.6), with CPU fallback for compatibility. Reproducibility was ensured by fixing random seed (SEED=42) across all stochastic operations.

## 3. Results and discussions

### 3.1. Model performance under five-fold cross-validation

We performed five-fold cross-validation on the final refined PN dataset that comprised a total of 1951 mutations from protein-DNA, protein-RNA, and protein-DNA-RNA hybrid complexes. Using 8:2 train-test split, we maintained 1560 mutations in training and 391 mutations in testing for each fold. The use of five-fold cross-validation enhanced the robustness and reliability of our model's performance across the extensive $\varDelta\varDelta G$ range (−4 to +8 kcal/mol). **Table 2** and **Figure 3b** present the final performance results, where the model achieved an average Pearson correlation coefficient (PCC) of 0.76, mean absolute error (MAE) of 0.52 kcal/mol, and root mean square error (RMSE) of 0.66 kcal/mol on the test set. Z-score outlier detector (threshold =1.5) helped us to overcome nearly about 10% outliers in our dataset. The highest observed performance reached a PCC of 0.80 with an MAE of 0.54 kcal/mol. Including outliers we got an average PCC of 0.55, while the average MAE and RMSE were 0.70 kcal/mol and 1.04 kcal/mol respectively (**Figure S1a**) in five-fold cross-validation. DeepPNI achieved its best performance after excluding tail-end outliers, which were primarily located in the highly stabilizing and destabilizing $\varDelta\varDelta G$ regions. Nevertheless, the model successfully captured these extreme cases as well, demonstrating its capability to generalize across the full spectrum of protein-nucleic acid binding free energy changes upon mutation.

**Table 2. Performance of DeepPNI on PN dataset.**

| Metric | 5-fold cross validation |
|---|---|
| PCC | 0.76 ± 0.03 |
| MAE | 0.52 ± 0.03 |
| RMSE | 0.65 ± 0.03 |

### 3.2. Prediction of mutational effects in protein-DNA and protein-RNA complexes

Furthermore, we evaluated the model's performance separately on the protein-DNA (PD) and protein-RNA (PR) datasets via 5-fold cross validation to examine its generalization across different complex types. The model achieved an average PCC of 0.75 for both PD and PR datasets (**Figure 3c** and **d**); however, the PR dataset exhibited slightly lower MAE (0.50 compared to 0.53 kcal/mol) and RMSE (0.64 compared to 0.67 kcal/mol) values compared to the PD dataset (**Table S1**), indicating more accurate predictions. Interestingly, the number of mutations in the protein-RNA complexes was approximately half that of the protein-DNA complexes. The $\varDelta\varDelta G$ range for both the datasets also varies drastically. We observed a range of -1.4 to +8 kcal/mol for the PR dataset and -4 to +6 kcal/mol PD dataset. We determined model's performance including the tail-ended $\varDelta\varDelta G$ values for protein-DNA and protein-RNA complexes separately. We observed an average PCC of 0.54 for PD and 0.52 for PR dataset in that case (**Figure S1b** and **S1c**). Among the five folds, we observed the highest PCC of 0.79 for PD and 0.85 for PR dataset without outliers. With these extreme outliers DeepPNI showed highest PCC of 0.60 and 0.68 for PD and PR dataset separately.

### 3.3. Model performance across temperature-based mutation groups

Experimental temperature is directly proportional to the binding free energy change upon mutation. Hence, we split the dataset in three groups, mutations @298K, mutations @>298K, and mutations @<298K. For these three groups, we performed five-fold cross-validation separately. While analysing the best performance, we observed that DeepPNI gave a better result for the mutations above 298K temperature with an average PCC value of 0.82 (**Figure 3e** and **Table S1**). Mutations at 298K temperature and below 298K temperature showed a similar average PCC of 0.77; however, for 298K temperature mutations the MAE and RMSE values were lower which indicates more stable and consistent predictions.

**Table 3. Performance of DeepPNI on external ProNAB data-split.** This data-split is based on experimental method for $\Delta\Delta G$ calculation.

| ProNAB data-split | PCC | MAE | RMSE |
|---|---|---|---|
| ITC | 0.62 | 0.60 | 0.74 |
| **GS** | **0.81** | **0.44** | **0.58** |
| **FP** | **0.81** | **0.50** | **0.67** |
| FB | 0.77 | 0.44 | 0.59 |

### 3.4 Ablation study

Our final model fused two feature components, RGCN and ESM-2. RGCN built embeddings using the atomic graph coming from the local structural environment of wild-type residue, and ESM-2 used the amino acid sequence around the wild-type amino acid. We performed ablation study to estimate importance of the features. Using RGCN and ESM-2 features separately, we got an average PCC of 0.52 and 0.73 in five-fold cross-validation, while concatenation of these two features gave an average PCC of 0.76 with a much lower MAE of 0.51 kcal/mol and RMSE of 0.64 kcal/mol. These results indicate the importance of both structural and sequential features for mutational binding free energy change prediction. We also compared the importance of relational features in this prediction task. We used simple GCN, where we kept all the edges between the nodes same, i.e., edges were not categorised here. We achieved an improvement of 2.70% after the introduction of relational feature (**Figure 3f** and **Table S2**) in the five-fold cross-validation.

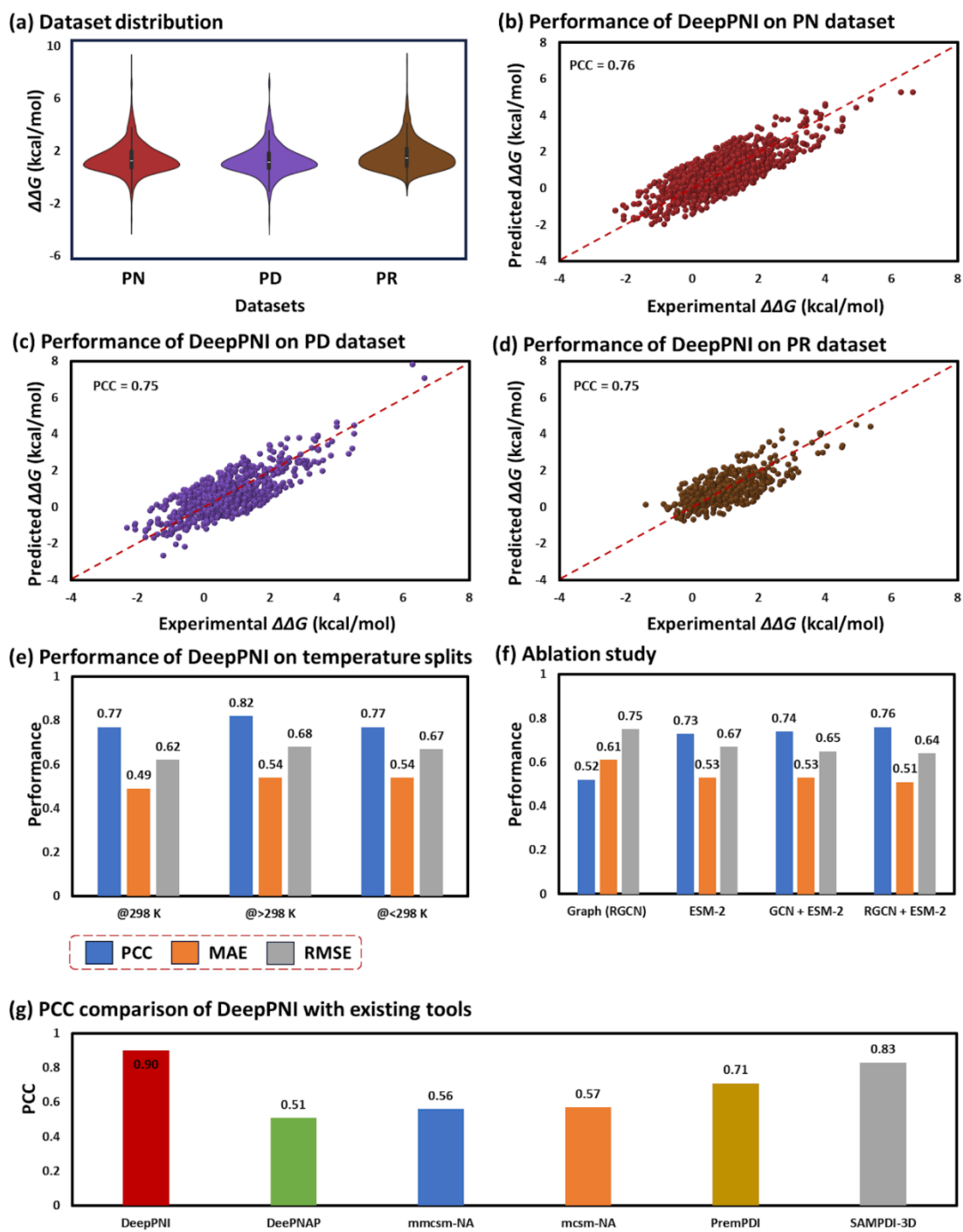

**Figure 3. Dataset and comparison results. (a)** Distribution *ΔΔG* of across PN, PD, and PR datasets**.** PN, PD, and PR contain 1951, 1328, and 620 mutations respectively. Scattered plot of performance of DeepPNI for **(b)** PN, **(c)** PD, and **(d)** PR dataset. **(e)** Performance of DeepPNI in different temperature ranges. **(f)** Ablation study result. **(g)** Comparison of DeepPNI with existing tools.

## 3.5. Evaluation using complex-based partitioning

Our PN dataset comprised of 439 unique PDB ID. We observed average of nearly four mutations per PDB ID. There were 107 PDB complexes that had single mutation, whereas1RTD, a protein-DNA complex had 30 mutations reported. We performed a complex-based training strategy to test the robustness of our model. We split the data in five-fold such that the training set contained mutations from 351 PDBs, and test set contained mutations from 88 unique PDBs. This splitting did not allow PDB complex overlap between training and test set. In this complex-based split, if one PDB complex mutation belonged from test set, no mutations from this PDB will be in training set. This split prevented data leakage problem.

DeepPNI showed a good performance across the splits with an average PCC of 0.47, while the MAE and RMSE values were 0.66 and 0.83 kcal/mol. Even in this type of complex-based split we got a good performance in terms of PCC, MAE, and RMSE. One of the folds achieved PCC of 0.54, with MAE and RMSE of 0.64 and 0.82 kcal/mol. This proved the robustness of DeepPNI (**Figure S2a, b** and **Table S3**).

## 3.6. Feature contribution analysis

The structural information of the protein-nucleic acid complexes was extracted via edge-aware RGCN. Our heterogeneous graph consists of three different types of nodes and six different types of edges. We took L2 normalization of the embedding vectors of three different types of nodes from the final RGCN layer. Nodes belonging to nucleic acid, wild-type residue, and other protein residue showed L2 normalization value of 0.82, 0.72 and 0.31 respectively (**Figure S3a**). It was clear that the interacting nucleic acid atoms and wild-type residue atoms played a key role in the mutational binding free energy prediction.

For capturing the node interaction importance in the $\Delta\Delta G$ prediction, we took an average of edge weights for different types of edges across the four RGCN layers. We found equal importance for all the edges (**Figure S3b**). But edge between wild-type residue atom and nucleic acid atom ($e_{wn}$) showed the highest average edge weight of 0.53 supporting the importance of nucleic acid atoms and wild-type residue atoms in the structural feature.

### 3.7. External validation

We built PN dataset from NABE database and further we splitted our data in PD, PR, and temperature-based splits. To test DeepPNI on unseen data, we collected data from ProNAB database and splitted them based on the experimental methodology. Among all groups, fluorescence polarization (FP), filter binding (FB), gel shift (GS), and isothermal titration calorimetry (ITC) had the maximum numbers of mutations. ITC, GS, FB, and FP had 210, 181, 173, and 117 mutations respectively.

We trained DeepPNI on our PN dataset, and used ITC, GS, FB, and FP sets as test sets. DeepPNI showed excellent results across these test sets (**Table 3**). Highest PCC value (0.81) was achieved by FP and GS followed by, FB (0.77) and ITC (0.62). These results proved our model's predictive power on external datasets.

### 3.8. Comparison with existing methods

To compare the prediction capability of DeepPNI, we used two methodologies. We compared DeepPNI's performance with two recent baseline models, DeePNAP and MutPNI. We also compared our model's performance with some well-known tools.

**Table 4. Comparison of DeepPNI with existing methods.**

| Dataset | Models | PCC | MAE | RMSE |
|---|---|---|---|---|
| PN | DeepPNI | **0.76** | **0.51** | **0.64** |
|  | DeePNAP | 0.52 | 0.70 | 1.01 |
| MPR311 | DeepPNI | **0.79** | **0.59** | **0.75** |
|  | MutPNI | 0.75 | 0.97 | 1.30 |
|  | DeepNAP | 0.59 | 0.82 | 1.15 |

We used PN dataset to compare DeepPNI with DeePNAP in five-fold cross-validation. DeepPNI outperformed DeePNAP (**Table 4**) with a PCC value of 0.76. The mean absolute error for DeePNAP and DeepPNI was 0.70 and 0.51 kcal/mol respectively. We achieved a lower RMSE of 0.64 kcal/mol than DeePNAP (1.01 kcal/mol). Furthermore, we did a ten-fold cross-validation on MPR311 dataset to compare DeepPNI, MutPNI, and DeepNAP. Here also, DeepPNI outperformed MutPNI and DeepNAP with an average PCC value of 0.79. Average MAE and RMSE values for DeepPNI were 0.59 and 0.75 kcal/mol. Whereas MutPNI achieved a MAE of 0.97 kcal/mol and RMSE 1.30 kcal/mol. The PCC, MAE, and RMSE for DeepNAP was 0.59, 0.82, and 1.15 kcal/mol.

Further to make a fair comparison with the existing tools, we tested DeepPNI on PremPDI dataset that had 193 mutations in common with PN dataset. Here in this case, DeepPNI outperformed other existing tools (**Figure 3g** and **Table S4**). The PCC value for DeepPNI was 0.90. SAMPDI-3D showed the nearest PCC value (0.83) to DeepPNI and DeePNAP showed the lowest PCC value of 0.51. These results showed the best performance of DeepPNI that was trained on a large dataset of 1951 mutation.

**Table 9. Comparison of predicted and experimental $\Delta\Delta G$ values for representative mutations across different $\Delta\Delta G$ regimes and protein-DNA-RNA hybrid complex.**

| PDB_ID | Mutation | Chain | Experimental $\Delta\Delta G$ (kcal/mol) | Predicted $\Delta\Delta G$ (kcal/mol) |
|---|---|---|---|---|
| 3Q8L | R100A | A | 6.64 | 5.27 |
| 3Q8L | K93A | A | 6.28 | 5.25 |
| 4R8I | R24A | A | 5.36 | 4.88 |
| 4R8I | K19A | A | 4.94 | 4.42 |
| 5AWH | Y463A | B | 4.53 | 3.73 |
| 2G1P | N126A | B | 0 | 0.08 |
| 1K8W | R181K | A | 0 | 0.05 |
| 5E24 | V479A | E | 0 | -0.40 |
| 1SKM | V121A | A | 0 | 0.05 |
| 6IIQ | N76A | B | 0 | -0.02 |
| 2VWJ | F116A | A | -2.31 | -1.23 |
| 2VWK | F116A | A | -2.31 | -1.22 |
| 1QTM | I614K | A | -2.09 | -0.92 |
| 2VWJ | P90I | A | -2.07 | -1.53 |
| 2VWK | P90I | A | -2.07 | -1.73 |
| 4WB2 | V709A | B | -0.39 | 0.35 |
| 4WB2 | R721A | B | 0.34 | -0.20 |
| 4WB2 | S697A | B | 3.49 | 0.28 |

### 3.9. Case study analysis

We did a systematic study to test DeepPNI's predictive accuracy (**Figure 4 and Table 5**). We built four group of mutations and compared our predicted $\Delta\Delta G$ with the experimental ones. Our group covered the complete vast regime of $\Delta\Delta G$. After outlier removal, we comprehensively collected mutations from top five strongly destabilizing (more positive value), five no changes in stability (0 value) and top five stabilizing (more negative value) mutations. We also observed the predicted ΔΔ$G$ value of protein-DNA-RNA hybrid complex. For the PDB complex 3Q8L, arginine was mutated to alanine at 100 position of chain A. Its experimental ΔΔ$G$ value was 6.64 kcal/mol, while DeepPNI predicted it as 5.27 kcal/mol. For PDB complex 2GIP, 1K8W, 5E24, 1SKM, and 6IIQ the experimental ΔΔ$G$ was 0 kcal/mol for the mutations N126A, R181K, V479A, V121A, and N76A respectively. In the stabilizing ΔΔ$G$ region, we again got good results compared to experimental ΔΔ$G$. For the protein-DNA-RNA hybrid complex three mutations were reported. All these mutations involved alanine amino acid. The experimental ΔΔ$G$ values for V709A, R721A, and S697A were −0.39, 0.34, and 3.49 kcal/mol, respectively, while DeepPNI predicted them as 0.35, −0.20, and 0.28 kcal/mol, respectively. DeepPNI performed good across the vast regime of ΔΔ$G$, proving its applicability for predicting both stabilizing and destabilizing mutation effects for protein-nucleic acid complexes.

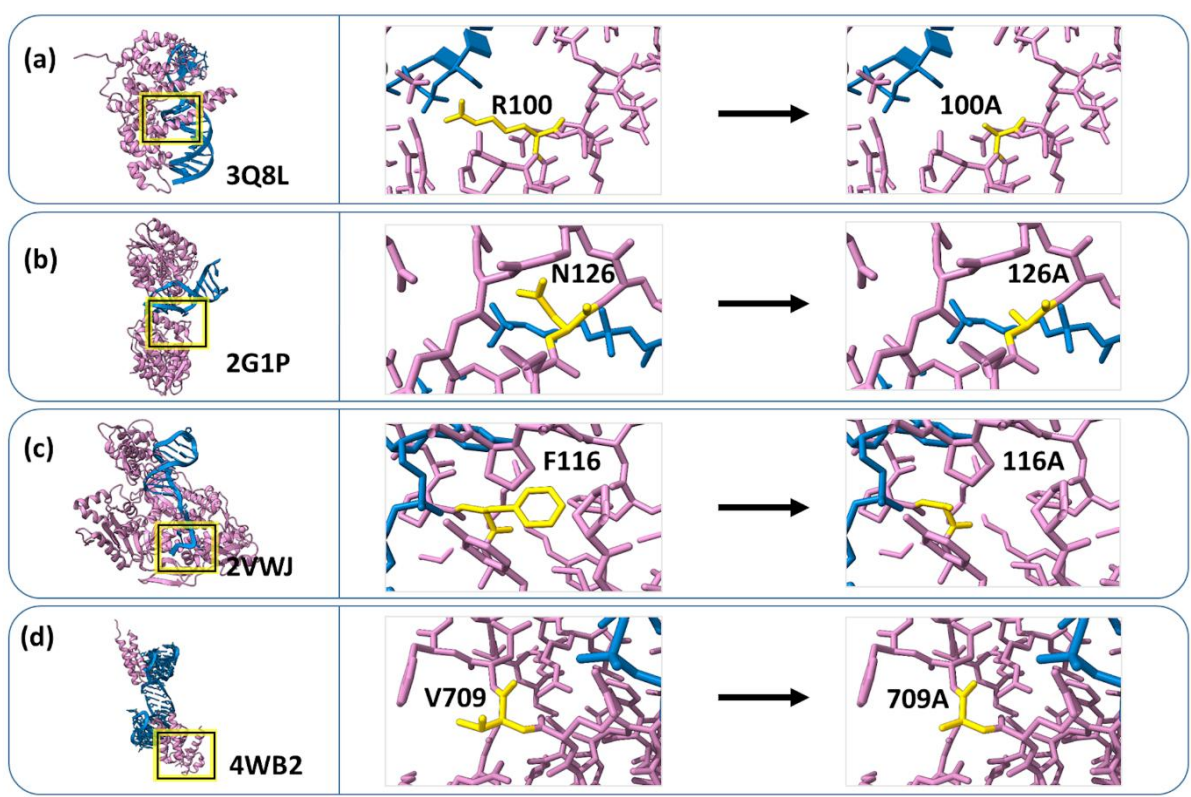

**Figure 4. Protein-nucleic acid complexes and their mutation site. (a)** PDB ID: 3Q8L, mutation: R100A; **(b)** PDB ID: 2GIP, mutation: N126A; **(c)** PDB ID: 2VWZ; mutation: F116A. (d) PDB ID: 4WB2; mutation: V709A.

# 4. Conclusions

In this study, we present DeepPNI, a deep learning framework that integrates graph-based structural representations with ESM-2 protein sequence embeddings for mutation-induced changes in binding free energy ($\Delta\Delta G$) prediction in protein-nucleic acid complexes. Comprehensive evaluation on the PN dataset comprising 1951 mutations demonstrates the superior predictive performance of DeepPNI, achieving a Pearson correlation coefficient (PCC) of 0.76 ± 0.03, mean absolute error (MAE) of 0.51 ± 0.03, and root mean square error (RMSE) of 0.64 ± 0.03 under five-fold cross-validation.

Further analysis across protein-DNA (PD) and protein-RNA (PR) complexes confirms the robustness and generalizability of DeepPNI, with a consistent average PCC value of 0.75 across the both datasets. Though for PR dataset the MAE and RMSE values are slightly better. The model also maintains stable performance under various thermal conditions, suggesting its applicability across a broad range of experimental scenarios. Ablation experiments revealed two major aspects: Firstly, the combination of graph-derived structural features and sequence-based embeddings is critical for achieving optimal predictive accuracy. Secondly, introduction of relational feature improves the prediction accuracy. From the feature analysis experiments, nucleic acid and wild-type amino acid residue are proved to be more important, while edges are contributing equally to the $\Delta\Delta G$ prediction in context of structural features.

Evaluation on the external ProNAB dataset, stratified by experimental methodologies (ITC, GS, FP, FB), further validates the model's generalization capability, with DeepPNI attaining the highest correlation (PCC = 0.81) for FP and GS measurements. Comparative analysis with existing computational predictors, including DeePNAP, mmCSM-NA, mcsm-NA, PremPDI, and SAMPDI-3D demonstrates that DeepPNI consistently outperforms current state-of-the-art approaches, achieving a maximum PCC of 0.90 across benchmark datasets.

Collectively, these findings make DeepPNI a robust, accurate, and generalizable predictor of mutation-induced binding free energy changes in protein-nucleic acid complexes. Its consistent performance across diverse datasets and experimental conditions highlights its potential as a valuable computational tool for mutation effect analysis of protein-nucleic acid complexes. Performance on a large dataset of 1951 mutations, and the incorporation of relational feature in the atomic heterogeneous protein-nucleic acid graph, make the approach innovative.

## ASSOCIATED CONTENT

**Supporting Information**

The supporting information contains analysis including ProNAB data split, split dataset results, ablation study results, complex-based partition results, literature review study, feature analysis, comparison with other existing tools, results with extreme outliers for PN, PD, and PR datasets.

# Supporting information

**Supplementary Section 1**

**ProNAB data split**

In ProNAB database experimental methodology was also listed for every mutation. A total of 173 mutations were determined by the filter binding (FB) assay, which evaluates binding affinity by measuring the retention of radio-labelled nucleic acids on a protein-bound filter. The gel shift (GS) method accounted for 181 mutations and detects binding interactions based on the mobility change of protein–nucleic acid complexes during electrophoresis. Fluorescence polarization (FP) experiments contributed 117 mutations, where changes in fluorescence anisotropy are used to assess protein–nucleic acid binding strength. A total of 210 mutations were measured using isothermal titration calorimetry (ITC), which directly quantifies binding thermodynamics through heat exchange upon complex formation.

**Table S1. Performance of DeepPNI on split datasets.**

| Split dataset | PCC | MAE | RMSE |
|---|---|---|---|
| PD | 0.75 | 0.53 | 0.67 |
| PR | 0.75 | 0.51 | 0.64 |
| @298 K | 0.77 | 0.49 | 0.62 |
| @>298 K | 0.82 | 0.54 | 0.68 |
| @<298 K | 0.77 | 0.54 | 0.67 |

**Table S2. Ablation study.**

| Feature | PCC | MAE | RMSE |
|---|---|---|---|
| Graph (RGCN) | 0.52 | 0.61 | 0.75 |
| ESM-2 | 0.73 | 0.53 | 0.67 |
| GCN + ESM-2 | 0.74 | 0.53 | 0.65 |
| RGCN + ESM-2 | 0.76 | 0.51 | 0.64 |

**Table S3. Performance of DeepPNI on complex-based split.**

| PCC | MAE | RMSE |
|---|---|---|
| 0.54 | 0.64 | 0.82 |

**Table S4. Comparison of DeepPNI with existing tools.**

| | DeepPNI | DeePNAP | mmcsm-NA | mcsm-NA | PremPDI | SAMPDI-3D |
|---|---|---|---|---|---|---|
| PCC | 0.90 | 0.51 | 0.56 | 0.57 | 0.71 | 0.83 |

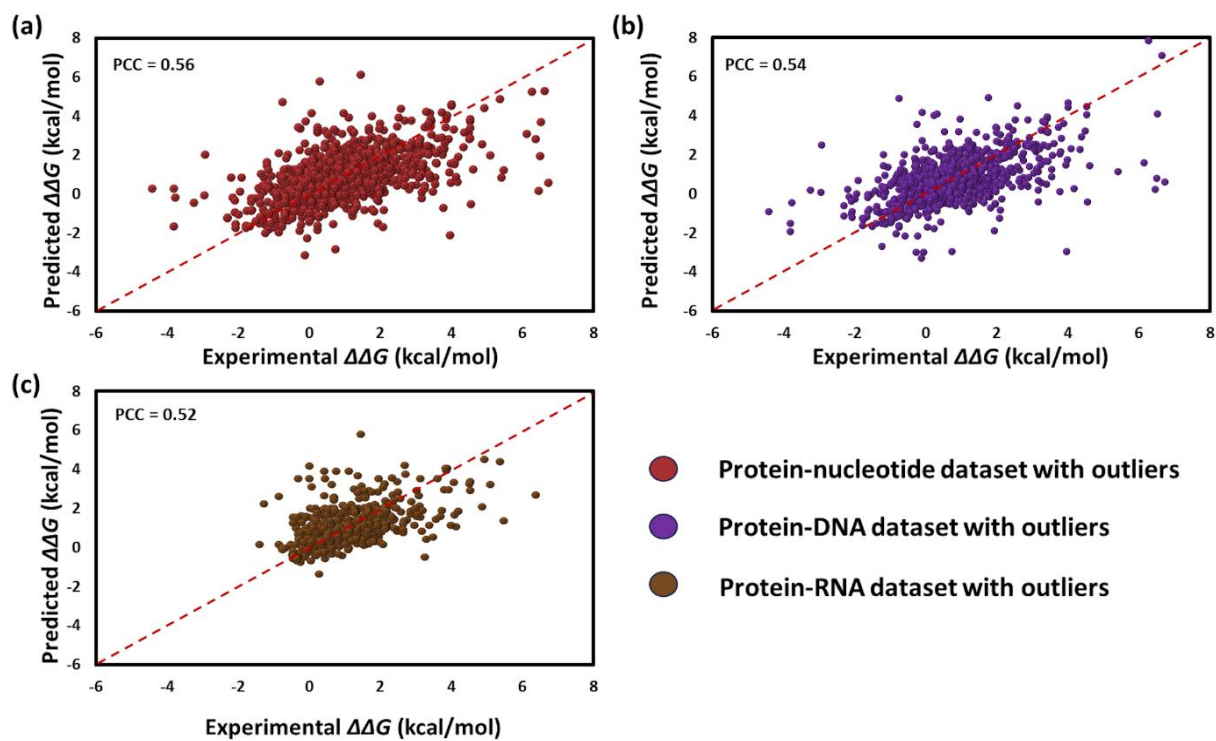

**Figure S1. Comparison results with outliers.** Scatter plot between experimental and predicted *ΔΔG* values for **(a)** PN, **(b)** PD, and **(c)** PR datasets with all the outliers.

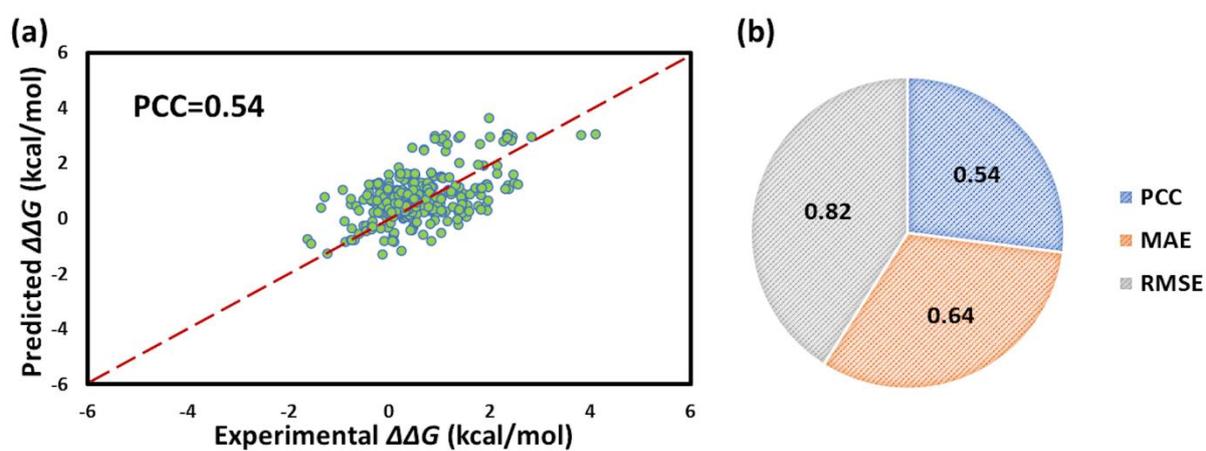

**Figure S2. Complex-based partition results. (a)** Scatter plot between experimental and predicted *ΔΔG*. **(b)** Pie chart of performance metrices PCC, MAE, and RMSE.

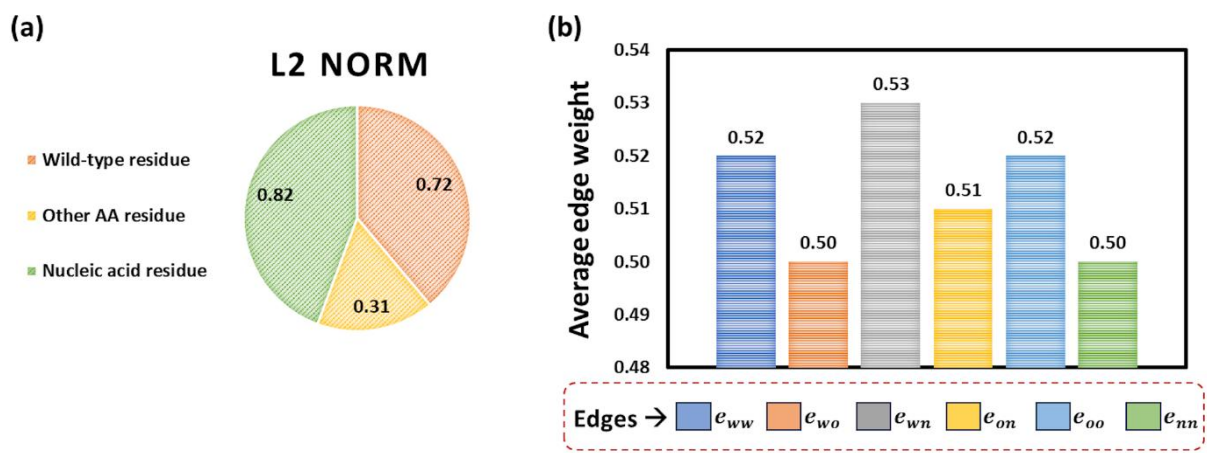

**Figure S3. Feature contribution analysis. (a)** L2 normalization of three different types of node embeddings at the final RGCN layer. **(b)** Average edge weights across the RGCN layers. Edge between atoms of wild-type residue ($e_{ww}$), edge between atoms of wild-type residue and other residue of protein ($e_{wo}$), edge between wild-type residue atom and nucleic acid atom ($e_{wn}$), edge between atoms of other residue ($e_{oo}$), edge between other residue and nucleic acid ($e_{on}$), edge between atoms of nucleic acid residue atom ($e_{nn}$).